\documentclass[%
preprint,showkeys,
 amsmath,amssymb,
 aps,
]{revtex4-1}

\usepackage{amsmath}
\usepackage{tabularx}
 \usepackage{graphics}
 \usepackage{epsfig}

\usepackage{amssymb}
\usepackage{graphicx}
\usepackage{multirow}
\usepackage[nodots,nocompress]{numcompress}
\newcolumntype{R}[1]{>{\raggedleft\arraybackslash}p{#1}}

\begin{document}



\title{Electron capture cross sections for stellar nucleosynthesis}



 \author{P.G. Giannaka}
 
  \email{pgiannak@cc.uoi.gr}
\author{T.S. Kosmas}%
 \email{hkosmas@uoi.gr}
 \affiliation{Division of Theoretical Physics, University of Ioannina, GR 45100 Ioannina, Greece.}
 
\begin{abstract}
In the first stage of this work, we perform detailed calculations for the cross sections of the electron capture on nuclei under laboratory conditions. Towards this aim we exploit the advantages of a refined version of the proton-neutron quasi-particle random-phase approximation (pn-QRPA) and carry out state-by-state evaluations of the rates of exclusive processes that lead to any of the accessible transitions within the chosen model space. In the second stage of our present study, we translate the above mentioned $e^-$-capture cross sections to the stellar environment ones by inserting the temperature dependence through a Maxwell-Boltzmann distribution describing the stellar electron gas. As a concrete nuclear target we use the $^{66}Zn$ isotope, which belongs to the iron group nuclei and plays prominent role in stellar nucleosynthesis at core collapse supernovae environment.

\end{abstract}


\maketitle
\section{Introduction}
Weak interaction processes occuring in the presence of nuclei under stellar conditions  play crucial role in the late stages of the evolution of massive stars and in the presupernova stellar collapse \cite{Fuller-Fowler-82,Aufd-Fush-94,Bethe-90,Lang-Pin-2003}.
As it is known, the core of a massive star, at the end of its hydrostatic burning is stabilized by electron degenerecy pressure as long as its mass does not exceed an  appropriate mass (the Chandrasekhar mass limit, $M_{Ch}$) \cite{Dean-Lang-98,Kol-Lang-98,Bethe-90,Niu-Paar-11,Paar-Colo-09}. When the core mass exceeds $M_{Ch}$, electron degeneracy pressure cannot longer stabilize the center of the star and the collapse starts. In the early stage of collapse electrons are captured by nuclei in the iron group region \cite{Bethe-90,Paar-Colo-09}.

During the pre-supernova evolution of core collapse supenova, the Fermi energy (or equivalently the chemical potential) of the degenerate electron gas is sufficiently large to 
overcome the threshold energy $E_{thr}$ ($E_{thr}$ is given by negative Q values of the reactions involved in the interior of 
the stars) \cite{Nabi-Ti-07} and the nuclear matter in the stellar core is neutronized. This high Fermi energy of the degenerate electron gas leads to enormous $e^-$-capture on 
nuclei and reduces the electron to baryon ratio $Y_e$ \cite{Nabi-Ti-11, Nabi-12}.
In this way, the electron pressure is reduced and the energy as well as the entropy drop.
One of the important characteristics of the early pre-explosion evolution is the fact that electron capture on nuclei (specifically on nuclei of the pf shell) plays a key role \cite{Cole-Ander-12,Zhi-Lang-11}. 

In the early stage of collapse (for densities lower than a few $10^{10} gr\, cm^{-3}$), the electron chemical potential is of the same order of magnitude as the nuclear Q value, and the $e^{-}$ capture cross-sections are sensitive to the details of GT strength distributions in daughter nuclei. For this reason, some authors restrict the calculations only to the GT strength and evaluate $e^{-}$-capture rates on the basis of the GT transitions 
(at these densities, electrons are captured mostly on nuclei with mass number $A\leq 60$) \cite{Dean-Lang-98,Paar-Colo-09,Nabi-12,Nabi-Ti-07,Zhi-Lang-11,Sar-Guer-Rodr-03}.
Various methods, used for calculating $e^-$-capture on nuclei during the collapse phase, have shown that this process produces neutrinos with rather low energies in contrast to the inelastic neutrino-nucleus reactions 
occuring in supernova \cite{Toi-Kol-01,Kol-Lang-03,Fr-Pin-06,Juo-lank-05}. These neutrinos escape the star carrying away energy and entropy from the core
which is an effective cooling mechanism of the exploding massive star \cite{Lang-Pin-03}.
For higher densities and tempratures, $e^{-}$ capture occurs on heavier nuclei $A\geq 65$ \cite{Paar-Colo-09,Niu-Paar-11,Nabi-Ti-11,Cole-Ander-12,Zhi-Lang-11}.
As a consequence, the nuclear composition is shifted to more neutron-rich and heavier nuclei (including 
those with $N>40$) which dominate the matter composition for densities larger than about $10^{10} gr\, cm^{-3}$ \cite{Lang-Pin-2003, Zhi-Lang-11,Lang-Kol-Dean-01,Lang-Pin-03}. 
 
The first calculations of stellar electron capture rates for iron group nuclei have been performed by employing the Independent Particle Model (IPM) \cite{Fuller-Fowler-82}. Recently, similar studies have been addressed by using Continuum RPA (CRPA) \cite{Kol-Lang-Vog-97}, large scale shell model \cite{Lang-Pin-00,Lang-Pin-Samp-01}, RPA \cite{Nabi-Ti-07}, etc \cite{Hix-03}. In the present work $e^-$-capture cross sections are obtained within a refined version of the Quasi-Particle  
Random Phase Approximation (QRPA) which is reliable for constructing all the accessible final 
(excited) states of the daughter nuclei in the iron group region of the periodic table \cite{Ring-Schuck,Kos-Ose-96,Has4,ts-kos-11,tsak-kos-11,kos-ts-12,Bal-Ydr-11,Balasi-Ydr-11,Bal-Ydr-12,Gian-Kos-13}. For the description of the required correlated nuclear ground states we determine single-particle 
occupation numbers calculated within the BCS theory \cite{Ring-Schuck,Has4,Gian-Kos-13}.  Our nuclear method is tested through the reproducibility of experimental muon capture rates relying on detailed calculations of exclusive, partial and total muon capture rates \cite{Kosm_new-limits,Eram-Kuz-Tet-98,Kol-Lang-00,Kosmas-01,Zin-Lang-06,Mark-Paar-09}. The agreement with experimental data provided us with high confidence level of our method and we continued with the calculations of electron capture cross sections in supernova conditions (where the densities and temperatures are high) using the pn-QRPA method.
In this paper,  we performed calculations for $^{66}Zn$ isotope (it belongs to the iron group nuclei) that plays prominent role in core collapse supernovae stellar nucleosynthesis \cite{Mayer-94,Kol-Lang-03,Fr-Pin-06}.

Our strategy in this work is, at first to perform extensive calculations of the transition rates for all the above mentioned nuclear processes assuming laboratory conditions, and then to translate these rates to the corresponding quantities within stellar environment through the use of an appropriate convolution procedure \cite{Dean-Lang-98,Lang-Pin-00,Lang-Pin-03,Cole-Ander-12,Zhi-Lang-11}. To this purpose, we assume that leptons under such conditions follow Maxwell-Boltzmann energy destribution  \cite{Dean-Lang-98,Lang-Pin-00}.

\section{Construction of nuclear ground and excited states}
\label{BCS-QRPA}
Electrons of energy $E_e$ are captured by nuclei interacting weakly with them via $W^{-}$ boson exchange as 
\begin{eqnarray}
\label{ecap}
(A,Z)+e^{-}\rightarrow (A,Z-1)^{*}+\nu_{e}
\end{eqnarray}
The outgoing $\nu_{e}$ neutrino carries energy $E_\nu$ while the daughter nucleus $(A, Z - 1)$ absorbs a part of the incident electron energy given by the difference between the initial $E_{i}$ and the final $E_{f}$ nuclear energies as $E_{\nu} = E_f - E_i$.

The nuclear calculations for the cross sections of the reaction (\ref{ecap}) start by
writing down the weak interaction Hamiltonian
$\hat{\mathcal{H}}_{w}$ which is given as a product of the leptonic, $j_{\mu}^{lept}$, and the hadronic, $\hat{\mathcal{J}}^{\mu}$, currents (current-current interaction Hamiltonian) as

\begin{eqnarray}  
\hat{\mathcal{H}}_{w}=
\frac{G}{\sqrt{2}}j_{\mu}^{lept}\hat{\mathcal{J}}^{\mu}
\end{eqnarray}
where $G=G_{F}cos\theta_{c}$ with $G_{F}$ and $\theta_{c}$ being the well known weak interaction coupling constant and the Cabbibo angle, respectively \cite{Has4,Gian-Kos-13,DonPe}.

From the nuclear theory point of view, the main task is to calculate the cross sections of the reaction (\ref{ecap}) which are based on the evaluation of the nuclear transition matrix elements between the initial  $\vert i \rangle$  and a final $\vert f \rangle$ nuclear states of the form
 
\begin{eqnarray}
\label{nucl-tran-ME}
\langle f|\widehat{H_{w}}|i\rangle = \frac{G}{\sqrt{2}} \, \ell^{\mu} \int d^{3}x \, e^{-i\textbf{q}\cdot\textbf{x}} \langle f|\widehat{\mathcal{J}_{\mu}}|i\rangle.
\end{eqnarray}
The quantity $\ell^{\mu} e^{-i\textbf{q}\cdot\textbf{x}}$ stands for the leptonic matrix element written in coordinate space with $\textbf{q}$ being the 3-momentum transfer.
For the  calculation of these transition matrix elements one may take advantage of the Donnelly-Walecka  multipole decomposition which leads to a set of eight independent irreducible tensor multipole operators containing polar-vector and axial-vector components \cite{DonPe} (see Appendix \ref{Nuclear Matrix Elements}).

In the present work, in Eq. (\ref{nucl-tran-ME}) the ground state of the parent nucleus $|i\rangle$ is computed by solving the relevant BCS equations which give us the quasi-particle energies and the amplitudes V and U that determine the probability for each single particle level to be occupied
or unoccupied, respectively \cite{Has4}.
Towards this aim, at first, we consider a Coulomb corrected Woods-Saxon potential with a spin orbit part as a mean field for the description of the strong nuclear field \cite{Tan-Pet-79,Bug-Bisn-79}. For the latter potential we adopt the parametrization of IOWA group \cite{IOWA}.
Then, we use as pairing interaction the monopole part of the Bonn C-D one meson exchange potential. The renormalization of this interaction to fit in the $^{66}Zn$ isotope, is achieved through the two pairing parameters $g_{pair}^{p,n}$, p (n) for proton (neutron) pairs, the values of which are tabulated in Table \ref{BCS}.

\begin{table}[h]
\caption{\label{BCS}Parameters for the renormalization of the interaction of proton
pairs, $g_{pair}^{p}$, and neutron pairs, $g_{pair}^{n}$. They have been
fixed in such a way that the corresponding experimental gaps,
$\Delta_{p}^{exp}$ and $\Delta_{n}^{exp}$ of $^{66}Zn$ isotope, are quite accurately reproduced.}
\begin{center}
\begin{tabular}{c|c|c|c|c|c|c}
\hline \hline\\[-0.75cm]
Nucleus &$g_{pair}^{n}$& $g_{pair}^{p}$  & $\Delta_{n}^{exp}\,(MeV)$  & $\Delta_{n}^{theor}\,(MeV)$  &$\Delta_{p}^{exp} \,(MeV)$ & $ \Delta_{p}^{theor}\,(MeV)$ \\[0.5ex]
\hline\\[-0.75cm]
    $^{66}Zn$   & 1.0059 & 0.9271  & 1.7715 & 1.7716 & 1.2815 & 1.2814 \\[0.5ex]
\hline \hline
\end{tabular}
\end{center}
\end{table}


As it is well known, the pairing parameters $g_{pair}^{p,n}$, are determined through the reproduction of the energy gaps, $\Delta_{p,n}^{exp}$, from neighboring nuclei as (3-point formula) 
\begin{eqnarray}
\Delta^{exp}_{n} =-\frac{1}{4}\Big[S_{n}(A-1,Z)-2S_{n}(A,Z)+ S_{n}(A+1,Z)\Big]
\end{eqnarray}
\begin{eqnarray}
\Delta^{exp}_{p} =-\frac{1}{4}\Big[S_{p}(A-1,Z-1)-2S_{p}(A,Z)+ S_{p}(A+1,Z+1)\Big]
\end{eqnarray}
where $S_{p}$ and $S_{n}$ are the experimental separation energies for protons and neutrons, respectively, of the target nucleus (A,Z) and the neighboring nuclei $(A\pm 1, Z\pm 1)$ and $(A\pm 1, Z)$. For the readers convenience in Table \ref{Sp-Sn} we show the values of experimental separation energies for the target $^{66}_{30}Zn$ and the neighboring nuclei $^{65}_{29}Cu$, $^{67}_{31}Ga$, $^{65}_{30}Zn$ and $^{67}_{30}Zn$.

\begin{table}[h]
\caption{The experimental separation energies in MeV for protons and neutrons of the target (A,Z) and neighboring  $(A\pm 1, Z\pm 1)$ and $(A\pm 1, Z)$ nuclei.}
\label{Sp-Sn}
\begin{small}
\begin{center}
\begin{tabular}{c|c|c|c|c|c|c}
\hline \hline\\[-0.75cm]
Nucleus & $S_{n}(A-1,Z)$ & $S_{n}(A,Z)$ & $S_{n}(A+1,Z)$ & $S_{p}(A-1,Z-1)$  & $S_{p}(A,Z)$  &$S_{p}(A+1,Z+1)$\\[0.5ex]
\hline\\[-0.75cm]
    $^{66}Zn$   &  7.979 & 11.059 & 7.052 & 7.454 &  8.924 & 5.269  \\[0.5ex]
\hline \hline
\end{tabular}
\end{center}
\end{small}
\end{table}


Subsequently, the excited states $|f\rangle$ of the studied daughter nucleus $^{66}Cu$ are  constructed
by solving the pn-QRPA equations \cite{Ring-Schuck,Kos-Ose-96,Has4,ts-kos-11,tsak-kos-11,kos-ts-12,Bal-Ydr-11,Balasi-Ydr-11,Bal-Ydr-12,Gian-Kos-13}, which in matrix form are written as \cite{Ring-Schuck}

\begin{eqnarray}\label{matr-form-QRPA}
\left(\begin{array}{cc}
        \mathcal{A} & \mathcal{B}\\
        \mathcal{-B} & \mathcal{-A}
          \end{array}\right)
\left(\begin{array}{c}
        X^{\nu} \\
        Y^{\nu} 
          \end{array}\right)=\Omega^{\nu}_{J^{\pi}}\left(\begin{array}{c}
        X^{\nu} \\
        Y^{\nu} 
          \end{array}\right).
\end{eqnarray}
$\Omega^{\nu}_{J^{\pi}}$ denotes the excitation energy of the QRPA state $|J^{\pi}_{\nu}\rangle$ with spin J and parity $\pi$. 

The solution of Eqs. (\ref{matr-form-QRPA}) is an eigenvalue problem which provides the amplitudes for forward and backward scattering X and
Y, respectively, as
well as the QRPA excitation energies $\Omega^{\nu}_{J^{\pi}}$ \cite{Has4,Bal-Ydr-11,Gian-Kos-13,ts-kos-11}. In our method the solution of the QRPA equations is
carried out separately for each multipole set of states
$|J^{\pi} \rangle $.

For the renormalization of the residual 2-body interaction (Bonn C-D potential), the strength parameters for the particle-particle $(g_{pp})$ and particle-hole $(g_{ph})$
interaction entering the QRPA matrices $\mathcal{A}$ and $\mathcal{B}$, are determined (separately  for each multipolarity) from the reproducibility of the low-lying experimental energy spectrum of the final nucleus.
The values of these parameters in the case of the spectrum of $^{66}Cu$ are listed in Table \ref{gpp-gph}.

\begin{table}[h!]
 \caption{Strength parameters for the
particle-particle $(g_{pp})$ and particle-hole $(g_{ph})$
interaction for various multipolarities (for the rest of multipolarities, $J^{\pi} \leq 5^{\pm}$, the bare 2-body interaction has been used), in the case of the spectrum of $^{66}Cu$ nucleus.}
\label{gpp-gph}
\begin{small}
\begin{center}
\begin{tabular}{l|ccccc||cccc}
\hline \hline
\multicolumn{6}{c ||}{Positive Parity States}  & \multicolumn{4}{c}{Negative Parity States}
\\
\hline\\[-0.75cm]
 $J^{\pi}$ & $0^{+}$ & $ 1^{+}$ & $2^{+}$ & $3^{+}$ & $4^{+}$  & $ 1^{-}$ & $2^{-}$ & $3^{-}$ & $4^{-}$ \\[0.6ex]

\hline
 $g_{pp}$ & 0.827 & 0.547 & 0.686 & 0.854 & 1.300  &  0.994 & 0.200 & 0.486 & 0.622  \\[0.6ex]
  $g_{ph}$ & 0.336 & 0.200 & 1.079 & 0.235 & 0.200 &  1.200 & 0.200 & 1.200 & 1.200  \\[0.6ex]
\hline \hline
\end{tabular}
\end{center}
\end{small}
\end{table}

\begin{figure}[h!]
\begin{center}
\includegraphics[scale = 1.0]{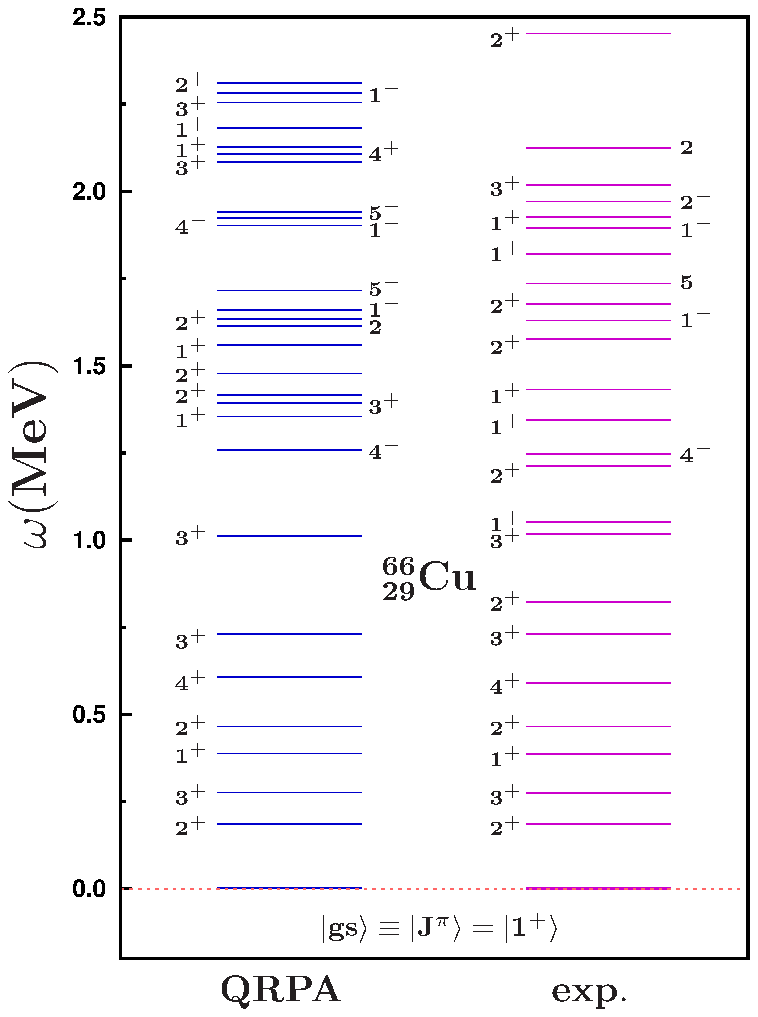}
\end{center}
\caption{Comparison of the theoretical
excitation spectrum (resulting from the solution of the QRPA
eigenvalue problem) with the low-lying (up to about 3 MeV)
experimental one for $^{66}Cu$ nucleus. As can be seen, the agreement is very good below 1 MeV but for higher excitation energies it becomes moderate.}
\label{Fasma}
\end{figure}

At this point, it is worth mentioning that for measuring the excitation energies of the daughter nucleus $^{66}Cu$ from the ground state of the initial one $^{66}Zn$, it is necessary a shifting of the entire set of QRPA eigenvalues. Such a shifting is required whenever in the pn-QRPA a BCS ground state is used, a treatment adopted by other groups previously \cite{Dean-Lang-98,Yous-Faes-09,Rod-Faes-06,Eram-Kuz-Tet-98}. The shifting for the spectrum of the daughter nucleus $^{66}Cu$, is done in such á way that  the first calculated value of each multipole state of $^{66}Cu$ (i.e. $1_{1}^{+}, 2_{1}^{+}$...etc.), to approach as close as possible the corresponding lowest experimental multipole excitation. 
Table \ref{shift} shows the shifting applied to our QRPA spectrum for each multipolarity of the parent nucleus $^{66}Zn$. 
We note that, a similar treatment is required in pn-QRPA calculations performed for double-beta decay studies where the excitations derived for the intermediate odd-odd nucleus (intermediate states) through p-n and n-p reactions from the neighboring nuclei, left or right nuclear isotope, do not match to each other \cite{Yous-Faes-09,Rod-Faes-06}.
\begin{table}[!t]
 \caption{The shift (in MeV) applied on the spectrum  (seperately of each multipole set of states) of $^{66}Cu$ isotope, daughter nucleus of the electron capture on $^{66}Zn$.}
 \vspace{0.2cm}
 \label{shift}
 \begin{center}
 \begin{tabular}{c|c||c|c}
 \hline
 \hline
 \multicolumn{2}{c ||}{Positive Parity States} & \multicolumn{2}{c}{Negative Parity States} \\[0.5ex]
 \hline
 $0^{+}$ & 0.90  & $0^{-}$  & 5.00  \\[0.5ex]
 \hline
 $1^{+}$ & 2.50  & $1^{-}$  & 6.80  \\[0.5ex]
 \hline
 $2^{+}$ & 2.55  & $2^{-}$  & 3.85  \\[0.5ex]
 \hline
 $3^{+}$ & 2.50  & $3^{-}$  & 2.60  \\[0.5ex]
 \hline
 $4^{+}$ & 1.75  & $4^{-}$  & 3.55  \\[0.5ex]
 \hline
 $5^{+}$ & 0.55  & $5^{-}$  & 3.00  \\[0.5ex]
 \hline \hline
 \end{tabular}
  \end{center}
 \end{table}
The resulting low-energy spectrum after using the parameters of Tables \ref{BCS} and \ref{gpp-gph} and the shifting shown in Table \ref{shift}, agrees well with the experimental one (see Fig. \ref{Fasma}). 
We must also mention that,  usually, in nuclear structure calculations we test a nuclear method in two phases:
First through the construction of the excitation spectrum as discussed before, and second through the calculations of electron scattering cross sections or muon capture rates.
Following the above steps, we test the reproducibility of the relevant experimental data for many nuclear models employed in nuclear applications (nuclear structure and nuclear reactions) and in nuclear astrophysics \cite{Lang-Pin-2003,Dean-Lang-98}.

\section{Results and Discussion}
\label{Results}

In this work we perform detailed cross section calculations for the electron capture on $^{66}Zn$ isotope on the basis of the pn-QRPA method.
The required nuclear matrix elements between the initial $|J_{i}\rangle$ and the final $|J_{f}\rangle$ states are determined by solving the BCS equations for the ground state \cite{Ring-Schuck,Has4,Gian-Kos-13} and the pn-QRPA equations for the excited states \cite{Has4,ts-kos-11,Bal-Ydr-11,Gian-Kos-13} (see Sect. \ref{BCS-QRPA}).
For the calculations of the original cross sections, a quenched value of $g_{A}$ (see Appendix \ref{Nuclear Form Factors}) is considered which subsequently modifies all relevant multipole matrix elements \cite{Zin-Lang-06,Mark-Paar-09,Hau-91,Wild-84}.

At this point of the present work and in order to increase the confidence level of our method, we perform total muon capture rates calculations \cite{Kosm_new-limits,Eram-Kuz-Tet-98,Kol-Lang-00,Kosmas-01,Zin-Lang-06,Mark-Paar-09}. The comparison with experimental and other theoretical results is shown in Sect. \ref{muon-capt-sect}. Afterwards, we study in detail the electron capture process as follows: i) Initially we consider laboratory conditions, i.e. the initial (parent) nucleus  is considered in the ground state and no temperature dependence is assumed (see Sect. \ref{Orig-Elec-Cap-CrSec-Sec}).
ii) Second, we consider stellar conditions, i.e. the parent nucleus is assumed to be in any initial excited state and due to the $e^{-}$-capture process it goes to any final excited state of the daughter nucleus. At these conditions it is necessary to take into account the temperature dependence of the cross sections (see Sec. \ref{St-El-Cap-CrSec-Sec}) \cite{Paar-Colo-09}.

\subsection{Calculations of Muon Capture Rates for $^{66}Zn$}
\label{muon-capt-sect}
Despite the fact that the muon capture on nuclei does not play a crucial role in stellar-nucleosynthesis, it is, however, important to start our study from this process since, the nuclear matrix elements required for an accurate description of the $\mu$-capture are the same for all semi-leptonic charge-changing weak interaction processes. In addition, the excitation spectrum of the daughter $(A, Z - 1)$ nucleus, as we saw before, is in good agreement with the experimental data. 

The calculations of the muon capture rates are performed in three steps:
In the first step we carry out realistic state-by-state calculations of exclusive Ordinary Muon Capture (OMC) rates in $^{66}Zn$ isotope for all multipolarities with $J^{\pi}\leq 5^{\pm}$ (higher multipolarities contribute negligibly). The appropriate expression for the exclusive muon capture rates is written as: 
 \begin{eqnarray}
 \label{OMC-rates}
\Lambda_{gs \rightarrow J_{f}^{\pi}} \equiv \Lambda_{J_{f}^{\pi}} = 2G^{2}\langle \Phi _{1s}  \rangle^{2} R_{f} q_{f}^{2} \Big[ \big|\langle J_{f} ^{\pi}\Vert (\widehat{\mathcal{M}}_{J}-\widehat{\mathcal{L}}_{J})\Vert 0_{gs}^{+}\rangle \big|^{2}
 +\big|\langle J_{f} ^{\pi}\Vert (\widehat{\mathcal{T}}_{J}^{el}-\widehat{\mathcal{T}}_{J}^{magn})\Vert 0_{gs}^{+}\rangle \big|^{2} \Big]
\end{eqnarray}
where $\Phi_{1s}$ represents the muon wave function in the 1s muonic orbit. The operators in Eq. (\ref{OMC-rates})  refer to as Coulomb $\mathcal{\widehat{M}}_{J}$, longitudinal $\mathcal{\widehat{L}}_{J}$, transverse electric $\mathcal{\widehat{T}}_{J}^{el}$ and transverse magnetic $\mathcal{\widehat{T}}_{J}^{mag}$ multipole operators (see  Appendix \ref{Nuclear Matrix Elements}). The factor $R_{f}$ in Eq. (\ref{OMC-rates}) takes into consideration the nuclear recoil which is written as 
$R_{f} = \Big( 1 + q_{f}/{M_{targ}}\Big)^{-1}$, 
with $M_{targ}$ being the mass of the target (parent) nucleus.   
 
Due to the fact that there are no available data in the literature for exclusive muon capture rates, the test of our method is realized by comparing partial and total muon capture rates with experimental data and other theoretical results \cite{Zin-Lang-06,Mark-Paar-09}. Towards this purpose,
our second step includes calculations of the partial $\mu^{-}$-capture rates for various low-spin multipolarities, $\Lambda_{J^{\pi}}$ (for $J^{\pi}\leq 4^{\pm}$), in the studied  nucleus. These partial rates are found by summing over the contibutions of all the individual multipole states of the studied multipolarity as
\begin{small}
\begin{eqnarray}
\Lambda_{J^{\pi}} = \sum_{f} \Lambda_{gs \rightarrow J_{f}^{\pi}} = 2G^{2}\langle \Phi _{1s}  \rangle^{2} &\Big[& \sum_{f} q_{f}^{2} R_{f} \big|\langle J_{f} ^{\pi}\Vert (\widehat{\mathcal{M}}_{J}-\widehat{\mathcal{L}}_{J})\Vert 0_{gs}^{+}\rangle \big|^{2}\nonumber\\
&+&\sum_{f} q_{f}^{2} R_{f}\big|\langle J_{f} ^{\pi}\Vert (\widehat{\mathcal{T}}_{J}^{el}-\widehat{\mathcal{T}}_{J}^{magn})\Vert 0_{gs}^{+}\rangle \big|^{2} \Big]
\end{eqnarray}
\end{small}
($f$ runs over all states of the multipolarity $\vert J^{\pi}\rangle$).
We also estimate the percentage (portion) of their contribution into the total $\mu$-capture rate for the most important multipolarities. In Table \ref{portions-of total mucap} we tabulate the individual portions of the low-spin multipole transitions ($ J^{\pi}=4^{\pm}$). As can be seen, the contribution of the $1^{-}$ multipole transitions is the most important multipolarity exhausting about $44\%$ of the total muon-capture rate.  Such an important contribution was found in $^{16}O$ and $^{48}Ca$ isotopes studied in Ref. \cite{Kol-Lang-00}.

\begin{table}[!t]
 \caption{\label{portions-of total mucap} The percentage of each multipolarity into the total muon-capture rate evaluated with our pn-QRPA method.}
 \begin{center}
\begin{tabular}{r| r|| r| r }
\hline \hline
\multicolumn{2}{c ||}{Positive Parity Transitions} & \multicolumn{2}{c}{Negative Parity Transitions}
\\
\hline\\[-0.75cm]
$J^{\pi}$ & Portions $(\%)$ & $J^{\pi}$ & Portions $(\%)$ \\
\hline
 $0^{+}$ & 8.22 & $0^{-}$ & 7.94 \\[0.2ex]
 \hline
 $1^{+}$ & 21.29 & $1^{-}$ & 44.21 \\[0.2ex]
 \hline
 $2^{+}$ & 2.85 & $2^{-}$ & 13.32  \\[0.2ex]
 \hline
 $3^{+}$ & 1.58 & $3^{-}$ & 0.34  \\[0.2ex]
 \hline
 $4^{+}$ & 0.01 & $4^{-}$ & 0.23  \\[0.2ex]
 \hline \hline
\end{tabular}
\end{center}
\end{table}

In the last step of testing our method, we evaluate total muon-capture rates for the $^{66}Zn$ isotope.
These rates are obtained by summing over all partial multipole transition rates (up to $J^{\pi} = 4^{\pm}$) as
\begin{eqnarray}
\Lambda_{tot} = \sum_{J^{\pi}} \Lambda_{J^{\pi}} = \sum_{J^{\pi}} \sum_{f} \Lambda_{J_{f}^{\pi}}
\end{eqnarray}

For the sake of comparison, the above mentioned $\mu$-capture calculations have been carried out using the quenched value $g_{A}= 1.135$ \cite{Zin-Lang-06,Mark-Paar-09}. The results are listed in Table \ref{Tot-mu-cap}, where we also include the experimental total rates as well as the theoretical ones of Refs. \cite{Zin-Lang-06} and \cite{Mark-Paar-09}. Moreover, in Table \ref{Tot-mu-cap} we show the individual contribution into the total muon capture rate of the polar-vector ($\Lambda^{V}_{tot}$), the axial-vector ($\Lambda^{A}_{tot}$), and the overlap ($\Lambda^{VA}_{tot}$) parts.
As can be seen, our results obtained with the quenched $g_{A}$ coupling constant are in very good agreement with the experimental total muon-capture rates (the deviations from the corresponding experimental rates are smaller than $7\%$). 
This agreement provides us with high confidence level for our method.

 \begin{table}[h]
 \caption{Individual contribution of Polar-vector, Axial-vector and Overlap parts into the total muon-capture rate. The total muon capture rates obtained by using the pn-QRPA with the quenched value of $g_{A}= 1.135$ for the medium-weight nucleus $^{66}Zn$, are compared with the available experimental data and with the theoretical rates of Ref. \cite{Zin-Lang-06} and Ref. \cite{Mark-Paar-09}.}
 \label{Tot-mu-cap}
 \begin{small}
\begin{center}
\begin{tabular}{c| c c c c |c |c c }
 \hline
 \hline
 \multicolumn{8}{c}{ Total Muon-capture rates $\Lambda_{tot} (\times 10^{6} s^{-1}$)} \\[0.5ex]

 \cline{1-8}\\[-0.75cm]
& \multicolumn{4}{c}{Present pn-QRPA Calculations} & \multicolumn{1}{|c|} {Experiment} & \multicolumn{2}{c}{Other theoretical Methods}\\[0.5ex]
 \cline{1-8}\\[-0.73cm]
   Nucleus &  $\Lambda^{V}_{tot}$  & $\Lambda^{A}_{tot}$ & $\Lambda^{VA}_{tot}$ & $\Lambda_{tot}$ & $\Lambda^{exp}_{tot}$ & $\Lambda^{theor}_{tot}$ \cite{Zin-Lang-06} & $\Lambda^{theor}_{tot}$ \cite{Mark-Paar-09}\\ [0.5ex]

  \hline \\[-0.73cm]
    $^{66}Zn$ & 1.651 & 4.487 & -0.204  & 5.934 & 5.809 & 4.976 & 5.809  \\[0.5ex]
\hline
\hline
\end{tabular}
\end{center}
\end{small}
\end{table}

\subsection{Electron Capture Cross Section}
After acquiring a high confidence level for our nuclear method, we proceed with the main goal of the present study which concerns the calculations of the electron-capture cross sections. As mentioned before, this includes original (see Section \ref{Orig-Elec-Cap-CrSec-Sec}) and stellar electron capture investigations (Section \ref{St-El-Cap-CrSec-Sec}).
 
\subsubsection{Original Electron Capture Cross Section on $^{66}Zn$ isotope}
\label{Orig-Elec-Cap-CrSec-Sec}

The original cross sections for the electron capture process in the $^{66}Zn$ isotope are obtained by using the pn-QRPA method considering all the accessible transitions of the final nucleus $^{66}Cu$.
In the Donnelly-Walecka formalism the expression for the differential cross section in electron capture by nuclei reads \cite{Paar-Colo-09}
\begin{eqnarray}
\label{EC-dif-CRSEC}
\frac{d\sigma_{ec}}{d\Omega} &=& \frac{G_{F}^{2} cos^{2}\theta_{c}}{2\pi} \frac{F(Z,E_{e})}{(2J_{i} +1)}\Big\lbrace \sum_{J \geq 1} \mathcal{W}(E_{e},E_{\nu}) \nonumber\\
&\times& \lbrace [ 1- \alpha cos\Phi + b sin^{2}\Phi] \big[\vert \langle J_{f}\Vert \mathcal{\widehat{T}}^{mag}_{J}\Vert J_{i}\rangle \vert ^{2}
+ \vert \langle J_{f}\Vert \mathcal{\widehat{T}}^{el}_{J}\Vert J_{i}\rangle \vert ^{2}\big] \nonumber\\
&-& \big[\frac{(\varepsilon_{i}+\varepsilon_{f})}{q}
(1-\alpha cos\Phi)-d \big] 
2Re\langle J_{f}\Vert \mathcal{\widehat{T}}^{mag}_{J}\Vert J_{i}\rangle \langle J_{f}\Vert \mathcal{\widehat{T}}^{el}_{J}\Vert J_{i}\rangle^{*} \rbrace \nonumber\\
&+&\sum_{J \geq 0 } \mathcal{W}(E_{e},E_{\nu})\lbrace  (1 + \alpha cos\Phi) \vert \langle J_{f}\Vert \mathcal{\widehat{M}}_{J}\Vert J_{i}\rangle \vert ^{2}\nonumber\\
 &+& (1 +\alpha cos\Phi - 2bsin^{2}\Phi) \vert \langle J_{f}\Vert \mathcal{\widehat{L}}_{J}\Vert J_{i}\rangle \vert ^{2} \nonumber\\
 &-& \big[\frac{\omega}{q} (1+ \alpha cos\Phi) + d\big]2Re \langle J_{f}\Vert \mathcal{\widehat{L}}_{J}\Vert J_{i}\rangle \langle J_{f}\Vert \mathcal{\widehat{M}}_{J}\Vert J_{i}\rangle^{*} \rbrace\Big\rbrace
\end{eqnarray}
where $F(Z,E_e)$ is the well known Fermi function \cite{Kol-Lang-03}. The factor $W(E_e, E_\nu) = E_{\nu}^{2}/{(1+E_{\nu}/M_{T})}$  accounts for the nuclear recoil \cite{Niu-Paar-11}, $M_{T}$ is the mass of the target nucleus and the parameters $\alpha$, b, d are given e.g. in Ref. \cite{Has4}. The nuclear transition matrix elements between the initial state $|J_{i}\rangle$ and a final state $|J_{f}\rangle$ correspond to the Coulomb $ \widehat{\mathcal{M}}_{JM}$, longitudinal $ \widehat{\mathcal{L}}_{JM}$, transverse electric $\widehat{\mathcal{T}}^{el}_{JM}$ and transverse magnetic $\widehat{\mathcal{T}}^{mag}_{JM}$ multipole operators (discussed in Appendix \ref{Nuclear Matrix Elements})

From the energy conservation in the reaction (\ref{ecap}), the energy of the outgoing neutrino $E_{\nu}$ is written as
\begin{eqnarray}
E_{\nu} = E_{e} -Q +E_{i} -E_{f}
\end{eqnarray}
which includes the difference between the initial $E_{i}$ and the final $E_{f}$ nuclear states. The Q value of the process is determined from the experimental masses of the parent ($M_{i}$) and the daughter ($M_{f}$) nuclei as $ Q = M_{f} - M_{i}$ \cite{Dean-Lang-98}.
 
It is worth mentioning that for low momentum transfer, various authors use the approximation $q\rightarrow 0$ for all multipole operators of Eq. (\ref{EC-dif-CRSEC}). Then, the transitions of the Gamow-Teller operator ($GT^{+} = \sum_{i} \tau^{+}_{i}\sigma_{i}$), provide the dominant contribution to the total cross section \cite{Dean-Lang-98}.
 
\begin{figure}[h!]
\begin{center}
\includegraphics[scale = 0.8]{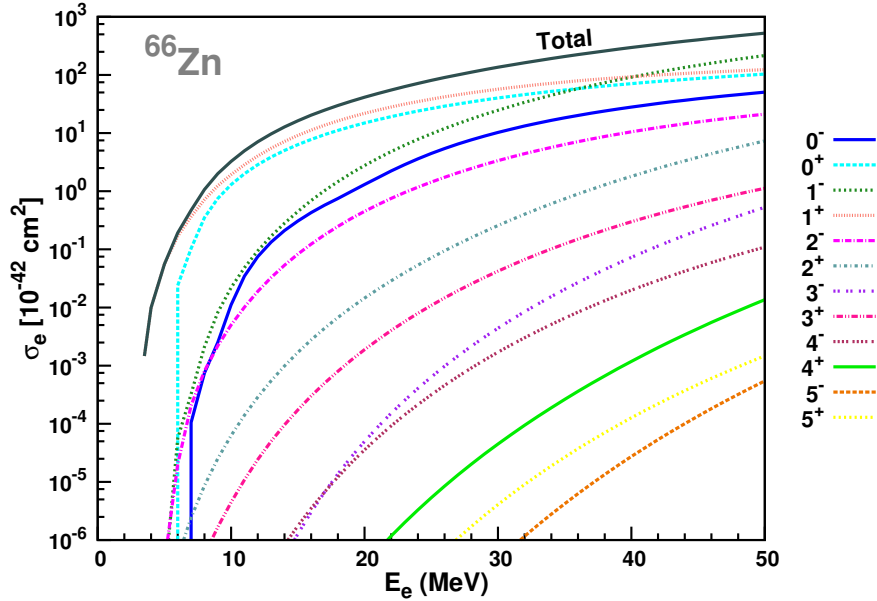}
\end{center}
\caption{Original total cross sections of electron-capture on the $^{66}Zn$ (parent) nucleus calculated with pn-QRPA method as a function of the incident electron energy $E_e$. The individual contributions of various multipole channels (for $J^{\pi}\leq 5^{\pm}$) are also demonstrated.}
\label{Zn-Orig-Cros-Sec}
\end{figure}

In performing detailed calculations for the original electron capture cross sections on $^{66}Zn$ isotope we assumed that (i) the initial state of the parent nucleus $^{66}Zn$ is the ground state $|0^{+}\rangle $ and (ii) the nuclear system is under laboratory conditions (no temperature dependence of the cross sections is needed).
The cross sections as a function of the incident electron energy $E_e$ are calculated with the use of realistic two-body interactions as mentioned before.
The obtained total original electron capture cross sections for $^{66}Zn$ target nucleus are illustrated in Fig. \ref{Zn-Orig-Cros-Sec} where the individual contributions of various multipole channels ($J^{\pi} \leq 5^{\pm}$) are also shown.
The electron capture cross sections in Fig. \ref{Zn-Orig-Cros-Sec} exhibit a sharp increase by several orders of magnitude within the first few MeV above energy-threshold, and this reflects the $GT^{+}$ strength distribution. For electron energy $E_e \geq 10 MeV$  the calculated cross sections show a moderate increase. 
From experimental and astrophysical point of view, the important range of the incident electron energy $E_e$ is up to 30 MeV. At these energies the $1^{+}$ multipolarity has the largest contribution to the total electron capture cross sections \cite{Dean-Lang-98,Paar-Colo-09}. In the present work we have extended  the range of $E_e$ up to 50 MeV since at higher energies (around 40 MeV) the contribution of other multipolarities like $1^{-}$, $0^{+}$ and $0^{-}$ become noticeable and can not be omitted (see Fig. \ref{Zn-Orig-Cros-Sec}). 

From the study of the original electron capture cross sections we conclude that, the total cross sections can be well approximated with the Gamow-Teller transitions only in the region of low energies \cite{Dean-Lang-98,Paar-Colo-09,Nabi-12,Nabi-Ti-07,Zhi-Lang-11,Sar-Guer-Rodr-03}. For higher incident electron energies the inclusion of the contributions originating from other multipolarities leads to better agreement \cite{Paar-Colo-09}.

\subsubsection{Stellar Electron Capture on $^{66}Zn$ isotope}
\label{St-El-Cap-CrSec-Sec}

As it is well known, electron capture process plays a crucial role in late stages of evolution of a massive star, in presupernova and in supernova phases \cite{Fuller-Fowler-82,Aufd-Fush-94,Bethe-90,Lang-Pin-2003}.
In presupernova collapse, i.e. at densities $\rho\leq 10^{10} g\,cm^{-3}$ and temperatures $300keV\leq T\leq 800keV$, electrons are captured by nuclei with $A\leq60$ \cite{Dean-Lang-98,Paar-Colo-09,Nabi-12,Nabi-Ti-07,Zhi-Lang-11,Sar-Guer-Rodr-03}.
During the collapse phase, at higher densities $\rho\geq 10^{10} g\,cm^{-3}$  and temperatures $T \simeq 1.0MeV$, electron capture process is carried out on heavier and more neutron rich nuclei with $Z<40$ and $N\geq40$ \cite{Paar-Colo-09,Niu-Paar-11,Nabi-Ti-11,Cole-Ander-12,Zhi-Lang-11}.

In an independent particle picture, the Gamow-Teller transitions (which is the most important in the electron capture cross section calculations) are forbidden for these nuclei \cite{Fuller-Fowler-82}. However, as  it has been demonstrated in several studies, GT transitions in these nuclei are unblocked by finite temperature excitations \cite{Lang-Kol-Dean-01,Lang-Pin-03}. At high temperatures, $T\simeq 1.5\,MeV$, GT transitions are thermally unblocked as a result of the excitation of neutrons from the pf-shell into the $g_{9/2}$ orbital.

For astrophysical environment, where the finite temperature and the matter density effects can not be ignored (the initial nucleus is at finite temperature), in general, the initial nuclear state needs to be a weighted sum over an appropriate energy distribution. Then, assuming Maxwell-Boltzmann distribution of the initial state $\vert i \rangle$ in Eq. (\ref{EC-dif-CRSEC}) \cite{Dean-Lang-98,Lang-Pin-00}, the total $e^{-}$-capture cross section is given by the expression \cite{Paar-Colo-09}
\begin{eqnarray}
\sigma(E_{e},T) &=& \frac{G_{F}^{2} cos^{2}\theta_{c}}{2\pi} \sum_{i} F(Z,E_{e})\frac{(2J_{i} + 1) e^{-E_{i}/(kT)}}{G(Z,A,T)} \nonumber\\
&\times&\sum_{f,J} (E_{e} -Q +E_{i} -E_{f})^{2}\frac{\vert \langle i\vert \widehat{O}_{J}\vert f\rangle\vert ^{2}}{(2J_{i} + 1)}
\end{eqnarray}
The sum over initial states in the latter equation denotes a thermal average of levels, with the corresponding partition function G(Z,A,T) \cite{Paar-Colo-09}. The finite temperature induces the thermal population of excited states in the parent nucleus. In the present work we assume that these excited states in the parent nucleus are all the possible states up to about 2.5 MeV. Calculations involving in addition other states lying at higher energies shows that they have no sizeable contribution to the total electron capture cross sections.
As mentioned before, for the evaluation of the  total electron capture cross sections, the use of a quenched value of $g_{A}$ is necessary \cite{Zin-Lang-06,Mark-Paar-09,Hau-91,Wild-84}. Since the form factor $F_{A}(q^{2})$ multiplies the four components of the axial-vector operator [see Eqs. (\ref{Coul}) - (\ref{Trans-Magn})], a quenched value of $g_{A}$ must enter the multipole operators generating the pronounced excitations $0^{-}, 1^{\pm}....$etc. For this reason, in our QRPA calculations we multiplied the free nucleon coupling constant $g_{A} = 1.262$ by the factor 0.8 \cite{Zin-Lang-06,Mark-Paar-09,Hau-91,Wild-84}.

\begin{figure}[h!]
\begin{center}
\includegraphics[scale = 0.8]{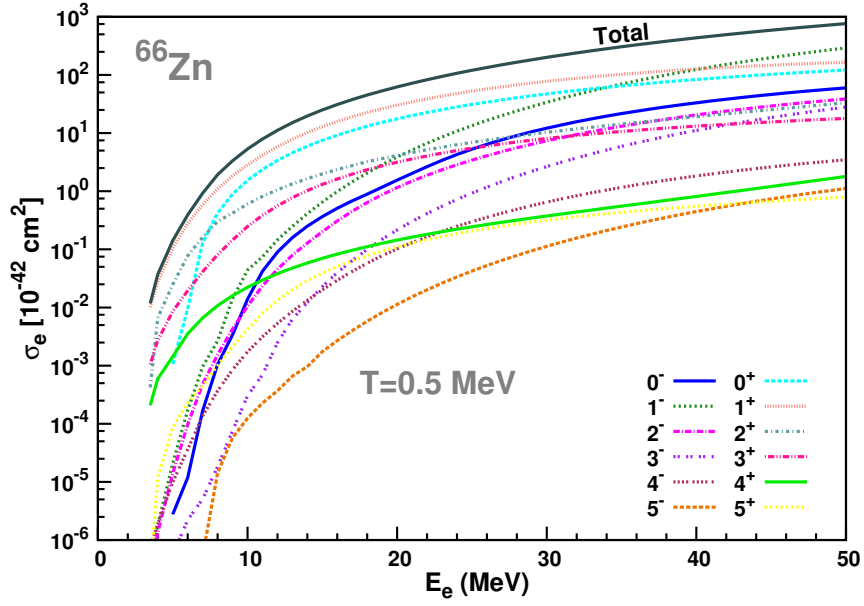}
\end{center}
\caption{Electron-capture cross sections for the $^{66}Zn$ parent nucleus at high temperature (T=0.5 MeV) in stellar environment obtained assuming Maxwell-Boltzmann statistics for the incident electrons. The total cross section and the dominant individual multipole channels ($J^{\pi}\leq 5^{\pm}$) are demonstrated as functions of the incident electron energy $E_e$.}
\label{Zn-Stellar-Cros-Sec}
\end{figure}

The results coming out of the study of electron capture cross sections under stellar conditions are shown in Fig. \ref{Zn-Stellar-Cros-Sec} where the same picture as in the original cross section calculations, but now with larger contribution is observed.
As discussed before, the dominant multipolarity is the $1^{+}$, which contributes more than $40\%$ into the total cross section. In the region of low energies (up to 30 MeV), the total $e^{-}$-capture cross section can be described by taking into account only the GT transitions, but at higher incident energies the contributions of other multipolarities become significant and can not be omitted. 

The percentage contributions of various multipolarities (with $J^{\pi} \leq 5^{\pm}$) into the total $e^{-}$-capture cross section at T=0.5 MeV and for incident electron energy $E_e =25MeV$ are tabulated in Table \ref{portions}. In addition, at this Table we list the values of the individual  $e^{-}$-capture cross sections of each multipolarity with $J^{\pi} \leq 5^{\pm}$. More specifically, for $E_e =25MeV$ the $1^{+}$ multipolarity contributes about $44\%$, the $0^{+}$ contributes about $26\%$ and the $1^{-}$ about $11\%$. The contributions coming from other multipolarities are less important (smaller than $5\%$).

\begin{table}[h!]
 \caption{Total $e^{-}$-capture cross sections (in $10^{-42}\, MeV^{-1}\,cm^{2}$) for $E_{e}=25MeV$. The percentage of each multipolarity into the total $e^{-}$-capture cross section  evaluated with our pn-QRPA code also tabulated here.}
\label{portions}
\begin{center}
\begin{tabular}{c|r|r||c|r|r}
\hline \hline
\multicolumn{3}{c ||}{Positive Parity Transitions} & \multicolumn{3}{c}{Negative Parity Transitions}
\\[0.5ex]
\hline\\[-0.75cm]
 $J^{\pi}$ & $\sigma_{e}(\times 10^{-42}\, \frac{cm^{2}}{MeV})$ & Portions $(\%)$ & $J^{\pi}$ & $\sigma_{e}(\times 10^{-42}\, \frac{cm^{2}}{MeV})$ & Portions $(\%)$  \\[0.7ex]
\hline
$0^{+}$ & 31.164 & 25.96 & $0^{-}$ & 5.288 & 4.41  \\[0.5ex]
$1^{+}$ & 52.779 & 43.98 & $1^{-}$ & 13.409 & 11.14 \\[0.5ex]
$2^{+}$ & 6.921 & 5.77  & $2^{-}$  & 3.262 & 2.72  \\[0.5ex]
$3^{+}$ & 5.499 & 4.58 & $3^{-}$  & 0.905 & 0.75  \\[0.5ex]
$4^{+}$ & 0.244 & 0.20  & $4^{-}$  & 0.299 & 0.25  \\[0.5ex]
$5^{+}$ & 0.208 & 0.17  & $5^{-}$  & 0.042 & 0.04  \\[0.5ex]
\hline \hline
\end{tabular}
\end{center}
\end{table}

\begin{figure}[h!]       
\begin{center}
\includegraphics[scale = 0.8]{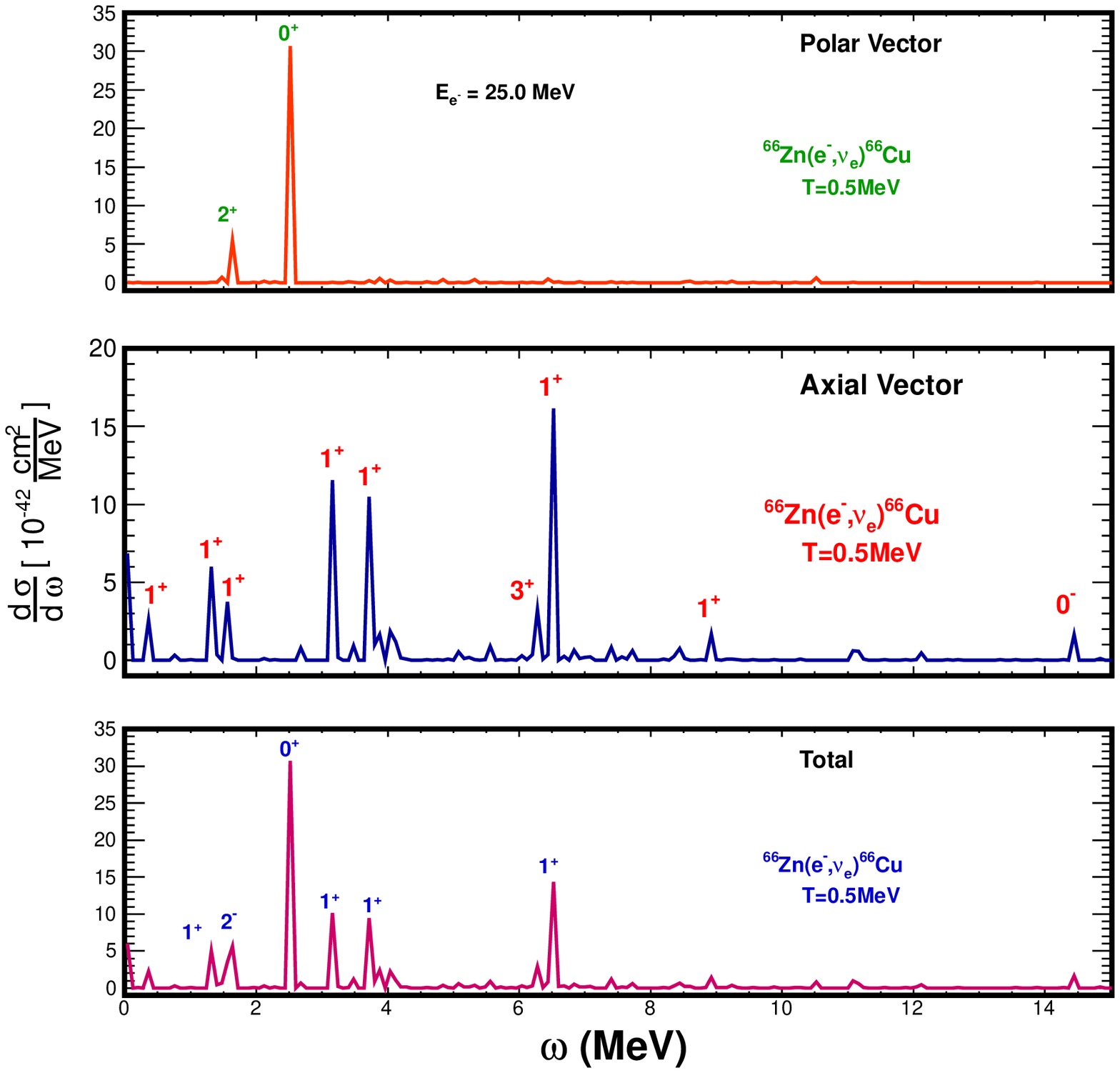}
\end{center}
\caption{Individual contributions of the Polar-Vector, ($\Lambda_{V}$), and Axial-Vector, ($\Lambda_{A}$), components as well as total electron-capture rate as functions of the excitation energy $\omega$  ($^{66}Zn$ is the parent nucleus).}
\label{Zn66-Sort-EC}
\end{figure}

In performing state-by-state calculations for the electron capture cross sections, our code has the possibility to provide separately the contribution of the polar-vector, the axial-vector and the overlap parts induced by the corresponding components of the electron capture operators.
In Fig. \ref{Zn66-Sort-EC} we illustrate the stellar differential cross sections of each individual transition of the polar-vector and axial-vector components. 

As mentioned before, our code gives separately the partial $e^{-}$-capture cross sections of each multipolarity. In order to study the dependence of the differential cross sections on the excitation energy $\omega$ throughout the entire pn-QRPA spectrum of the daughter nucleus, a rearrangement of all possible excitations $\omega$ in ascending order, with the corresponding cross sections, is required. This was performed by using a special code appropriate for matrices \cite{ts-kos-11}. In the model space chosen for $^{66}Zn$ isotope, for all multipolarities up to $J^{\pi}=5^{\pm}$ we have a number of 447 final states. The differential electron capture cross sections illustrated in Fig. \ref{Zn66-Sort-EC} present some characteristic clearly pronounced peaks at various excitation energies $\omega$. These peaks correspond mainly to $0^+,1^+$ and $2^{+}$ transitions.
More specifically, in the $^{66}Cu$ daughter nucleus the maximum peak correspond to the $0^{+}_1$ QRPA transition at $\omega = 2.538MeV$ and other characteristic peaks correspond to $1^{+}_7$,$1^{+}_8$ and $1^{+}_{10}$ transitions, located at energies $\omega = 3.194MeV$, $\omega = 3.686MeV$ and $\omega = 6.555MeV$, respectively (see Fig. \ref{Zn66-Sort-EC}). There are also shown other less important peaks in Fig. \ref{Zn66-Sort-EC}.

Before closing, it should be mentioned that, the e-capture cross sections presented in this work, may be useful in estimating neutrino-spectra arising from e-capture on nuclei during supernova phase. The knowledge of  $\mathcal{\nu}$-spectra at every point and time in the core is quite relevant for simulations of the final collapse and explosion phase of a massive star. As it is known \cite{Lang-Pin-03}, in the collapse phase, neutrinos are mainly produced by e-capture on nuclei and on free protons. The energy spectra of the emerging neutrinos from both reactions are important ingredients in stellar modelling and stellar simulations \cite{Lang-Pin-03,Lang-Pin-Samp-01}.

Furthermore, in core collapse simulations one defines the reaction rate of electron capture on nuclei given by 
\begin{eqnarray}
\label{reaction rate}
R_{h} = \sum_i Y_i \lambda_i
\end{eqnarray} 
where the sum runs over all nuclear isotopes present in the astrophysical environment ($Y_i$ denotes the abundance of a given nuclear isotope and $\lambda_i$ is the calculated electron capture rate for this isotope). The rates of Eq. (\ref{reaction rate}) must be known for a wide range of the parameters: T (temperature) and $\rho$ (nuclear density) of the studied star. 
Thus, for the calculation of the quantity $Y \cdot\lambda$ of a specific nuclear isotope one needs to know in addition to the nuclear composition Y the electron capture rates $\lambda$ calculated as we have shown in our present work.
The rates of electron capture on various nuclear isotopes and the corresponding emitted neutrino spectra  in the range of the parameters (T, $\rho$, $Y_e$) describing the star until reaching equilibrium during the core collapse, are comprehensively studied in Ref. \cite{Lang-Pin-03,Lang-Pin-00,Lang-Pin-Samp-01} for a great number of nuclear isotopes by using the large scale shell model. We are currently performing similar calculations for a set of isotopes by employing the present pn-QRPA method \cite{PG-14}.

Furthermore, the average neutrino energy, $\langle E_{\nu}\rangle$, of the neutrinos emitted by e-capture on nuclei can be obtained by dividing the neutrino-energy loss rate (defined by an expression similar to Eq. (\ref{reaction rate}) by replacing the rate $\lambda_i$ with the energy loss rate $E_j$) with the reaction rate for e-capture on nuclei $R_{h}$. 
Assuming e.g. power-law energy distribution for the neutrino spectrum produced by the e-capture in supernova phase, the average neutrino-energy $\langle E_{\nu}\rangle$ determines a specific  supernova-neutrino scenario.
In addition, the neutrino emissivity is obtained by multiplying the electron capture rate at nuclear statistical equilibrium with the neutrino-spectra \cite{Lang-Pin-03,Lang-Pin-00,Lang-Pin-Samp-01}.
Finally we note that, the rates for the inverse neutrino absorption process are also determined from the electron capture rates obtained as discussed in this section \cite{Lang-Pin-03}.

\section{Summary and Conclusions} 
The electron capture on nuclei plays crucial role during the presupernova and
collapse phase (in the late stage $e^{-}$-capture on free protons is also significant). It
becomes increasingly possible as the density in the star's center
is enhanced and it is accompanied by an increase of the chemical
potential (Fermi energy) of the degenerate electron gas. This process
reduces the electron-to-baryon ratio $Y_e$ of the matter
composition.

In this work, by using  our numerical approach based on a refinement of the pn-QRPA that describes reliably all the semi-leptonic weak interaction processes in nuclei,
we studied in detail the electron capture process on $^{66}Zn$ isotope and calculated original as well as stellar $e^{-}$-capture cross sections.
We tested our nuclear model (the pn-QRPA) through the reproducibility of orbital muon capture rates for this isotope.
The agreement with experimental data and other reliable theoretical results of partial and total $\mu$-capture rates as well as of the percentage contributions of various low-lying excitations is quite good which provides us with high confidence level for the obtained cross sections.

Our future plans are to extent the application of this method and make similar calculations for other interesting nuclei \cite{PG-14}. Also this method could be applied  to other semi-leptonic nuclear processes like beta-decay and charged-current neutrino-nucleus processes important in nuclear astrophysics and neutrino nucleosynthesis.

 \subsection*{Acknowledgments}
 This research has been co-financed by the European Union (European Social Fund-ESF) and Greek national funds through the Operational Program ``Education and Lifelong Learning" of the National Strategic Reference Framework (NSRF) - Research Funding Program: Heracleitus II. Investing in knowledge society through the European Social Fund.

\appendix
\section{Nuclear Matrix Elements}
 \label{Nuclear Matrix Elements}
The eight different tensor multipole operators entering  the above Equations (see Sec. \ref{Results}) refer to as Coulomb $\widehat{\mathcal{M}}_{JM}$, longitudinal $\widehat{\mathcal{L}}_{JM}$, transverse electric $\widehat{\mathcal{T}}^{el}_{JM}$ and transverse magnetic  $\widehat{\mathcal{T}}^{magn}_{JM}$, are defined as 

\begin{eqnarray}
\widehat{\mathcal{M}}_{JM}(qr) = \widehat{M}^{coul}_{JM} + \widehat{M}^{coul5}_{JM} \, , \quad \widehat{\mathcal{L}}_{JM}(qr) = \widehat{L}_{JM} + \widehat{L}^{5}_{JM}
\end{eqnarray}


\begin{eqnarray}
\widehat{\mathcal{T}}^{el}_{JM}(qr) = \widehat{T}^{el}_{JM} + \widehat{T}^{el5}_{JM} \, , \quad \widehat{\mathcal{T}}^{magn}_{JM}(qr) = \widehat{T}^{magn}_{JM} + \widehat{T}^{magn5}_{JM}
\end{eqnarray}


These multipole operators contain polar-vector as well as axial-vector parts and are written in terms of seven independent basic multipole operators as

\begin{eqnarray}
\label{Coul}
\hspace{-0.2cm}
\widehat M_{JM}^{coul}(q{\bf r})  = 
F_1^V(q_\mu^2)M^J_M(q{\bf r})
\end{eqnarray}

\begin{eqnarray}
\label{Long}
 \hspace{-0.2cm}
\widehat L_{JM}(q{\bf r})  = 
\frac{q_0}{q}\hat{M}_{JM}^{coul}(q{\bf r})
\end{eqnarray}

\begin{eqnarray}
\label{Trans-Electr}
\hspace{-0.2cm}
\widehat T_{JM}^{el}(q{\bf r})  =
\frac{q}{M_N}\left[F_1^V(q_\mu^2){\Delta ^{\prime}}^J_{M}(q{\bf
r}) +\frac{1}{2}\mu^V(q_\mu^2){\Sigma}^J_{M}(q{\bf
r})\right]
\end{eqnarray}

\begin{eqnarray}
\label{Trans-Magn}
\hspace{-0.2cm}
i\widehat T_{JM}^{mag}(q{\bf r})  = 
\frac{q}{M_N}\left[F_1^V(q_\mu^2)\Delta ^J_{M}(q{\bf r})
-\frac{1}{2}\mu^V(q_\mu^2){\Sigma^{\prime}}^J_{M}(q{\bf r})\right]
\end{eqnarray}

\begin{eqnarray}
\label{Coul5}
\hspace{-0.2cm}
i \widehat M_{JM}^5(q{\bf r})  = 
\frac{q}{M_N}\left[F_A(q_\mu^2)\Omega^J_{M}(q{\bf
r})+\frac{1}{2}(F_A(q_\mu^2) +q_0F_P(q_\mu^2)){\Sigma^
{\prime\prime}}^J_{M}(q{\bf r})\right]
\end{eqnarray}

\begin{eqnarray}
\label{Long5}
\hspace{-0.2cm}
-i\widehat L_{JM}^5(q{\bf r})  =
\left[F_A(q_\mu^2)-\frac{q^2}{2M_N}F_P(q_\mu^2)\right]{\Sigma^
{\prime\prime}}^J_{M}(q{\bf r})
\end{eqnarray}

\begin{eqnarray}
\label{Trans-Electr}
\hspace{-0.2cm}
-i\widehat T_{JM}^{el5}(q{\bf r}) 
=F_A(q_\mu^2){\Sigma^{\prime}}^J_{M}(q{\bf r})
\end{eqnarray}

\begin{eqnarray}
\label{Trans-Magn5}
 \hspace{-0.2cm}
\widehat T_{JM}^{mag5}(q{\bf r}) \, =
\,F_A(q_\mu^2){\Sigma}^J_{M}(q{\bf r})
\end{eqnarray}
where the form factors $F_{X}$, X=1,A,P and $\mu^{V}$ are functions of the 4-momentum transfer $q^{2}_{\mu}$ and $M_N$ is the nucleon mass.

These multipole operators, due to the Conserved Vector Current (CVC) theory, are reduced to seven new basic operators expressed in terms of spherical Bessel functions, spherical harmonics and vector spherical harmonics (see Refs. \cite{Mark-Paar-09,DonPe,Has4}). 
The single particle reduced matrix elements of the form $\langle j_{1} \Vert T_{i}^{J}\Vert j_{2}\rangle$, where $T_{i}^{J}$ represents any of the seven basic multipole operators ($M_{M}^{J}$, $\Omega_{M}^{J}$, $\Sigma_{M}^{J}$, $\Sigma_{M}^{'J}$, $\Sigma_{M}^{''J}$, $\Delta_{M}^{J}$, $\Delta_{M}^{'J}$) of Eq. (\ref{Coul})-(\ref{Trans-Magn5}), have been written in closed compact expressions as \cite{Has4}
\begin{eqnarray}
\langle (n_{1}l_{1})j_{1} \Vert T^{J}\Vert (n_{2}l_{2})j_{2}\rangle = e^{-y} y^{\beta/2}\sum_{\mu=0}^{n_{max}}P_{\mu}^{J}y^{\mu}
\end{eqnarray}
where the coefficients $P_{\mu}^{J}$ are given in Ref. \cite{Has4}.
In the latter summation the upper index $n_{max}$ represents the maximun harmonic oscillator quanta included in the active model space chosen as $n_{max} = (N_1+N_2-\beta)/2$, where $N_{i} = 2n_{i}+l_{i}$, i=1,2, and $\beta$  is related to the rank of the above operators \cite{Has4}.

In the context of the pn-QRPA, the required reduced nuclear matrix elements between the initial $\vert 0^{+}_{gs}\rangle$ 
and any final $\vert f\rangle$ state entering the rates of Eq. (\ref{OMC-rates}) are given by
\begin{eqnarray}
\label{red-matr-elem}
\langle f \Vert\widehat{T}^{J}\Vert 0^{+}_{gs} \rangle = \sum _{j_{2}\geq j_{1}} 
\frac{\langle j_{2}\Vert\widehat{T}^{J}\Vert j_{1} \rangle}{[J]} \,
\left[ X_{j_{2}j_{1}} u_{j_{2}}^{p} \upsilon_{j_{1}}^{n}
+ Y_{j_{2}j_{1}} \upsilon_{j_{2}}^{p} u_{j_{1}}^{n} \right]
\end{eqnarray}
where $u_j$ and $\upsilon_j$ are the probability amplitudes for the $j$-level to be unoccupied or occupied, respectively (see the text) \cite{Has4,Gian-Kos-13}.

These matrix elements enter the description of various semi-leptonic weak interaction processes in the presence of nuclei \cite{Has4,Gian-Kos-13,ts-kos-11,Bal-Ydr-11,Balasi-Ydr-11,tsak-kos-11,Bal-Ydr-12,kos-ts-12,Pap-Kos-14,Pap-Kos-13,DonPe}

\section{Nuclear Form Factors}
 \label{Nuclear Form Factors}
In Eqs. (\ref{Coul}) - (\ref{Trans-Magn5}) the standard set of free nucleon form factors $F_{X}(q^{2}_{\mu})$, X= 1, A, P and $\mu^{V}(q^{2}_{\mu})$ reads 
\begin{eqnarray}
F_{1}^{V}(q^{2}_{\mu}) = 1.000\Big[1+\big(\frac{q}{840\,MeV}\big)^{2}\Big]^{-2}
\end{eqnarray} 
\begin{eqnarray}
\mu^{V}(q^{2}_{\mu})= 4.706\Big[1+\big(\frac{q}{840\,MeV}\big)^{2}\Big]^{-2}
\end{eqnarray}  
\begin{eqnarray}
F_{A}(q^{2}_{\mu}) = g_{A}\Big[1+\big(\frac{q}{1032\,MeV}\big)^{2}\Big]^{-2}
\end{eqnarray}

\begin{eqnarray}
F_{P}(q^{2}_{\mu}) =\frac{ 2\,M_{N}\, F_{A}(q^{2}_{\mu})}{q^{2}+m_{\pi}^{2}}
\end{eqnarray}
where $M_{N}$ is the nucleon mass and $g_{A}$ is the axial vector free nucleon coupling constant (see the text).


\end{document}